\documentclass[12pt]{article}
\usepackage{amssymb}
\usepackage{graphicx}
\usepackage{setspace}

\newcommand{\n}{\nonumber \\}

\begin{document}
\begin{titlepage}
\begin{flushright}
KEK-TH-1849
\end{flushright}
\begin{center}

\vspace*{10mm}

{\LARGE\bf
\begin{spacing}{1.2}
Revolving D-branes and \\ Spontaneous Gauge Symmetry Breaking
\end{spacing}
}

\vspace*{20mm}

{\large
Satoshi Iso${}^{\; a,b}$ and Noriaki Kitazawa${}^{\; c}$
}
\vspace{10mm}

{${}^a$\sl\small KEK Theory Center, High Energy Accelerator Research Organization (KEK),\\ }
{${}^b$\sl\small Graduate University for Advanced Studies (SOKENDAI), \\ Tsukuba, Ibaraki 305-0801, Japan \\}
e-mail: {\small \it satoshi.iso(at)kek.jp}
\vspace{8pt}

{${}^c$\sl\small Department of Physics, Tokyo Metropolitan University, \\ Hachioji, Tokyo 192-0397, Japan \\ }
e-mail: {\small \it kitazawa(at)phys.se.tmu.ac.jp}
\vspace{8pt}


\begin{abstract}
\centerline{{\it Eppur si muove} -- Galileo Galilei, 1633}
\noindent
We propose a new mechanism of spontaneous gauge symmetry breaking 
 in the world-volume theory of revolving D-branes around a fixed point of orbifolds.
In this paper, we consider a simple model of the $T^6/{\bf Z}_3$ orbifold
 on which we put D3-branes, D7-branes and their anti-branes.
The configuration breaks supersymmetry,
 but the R-R tadpole cancellation conditions are satisfied.
A set of three D3-branes at an orbifold fixed point can separate from the point,
 but when they move perpendicular to the anti-D7-branes put on the fixed point,
 they are forced to be pulled back  due to an attractive interaction between 
 the D3 and anti-D7 branes. 
In order to stabilize the separation of the D3-branes at  nonzero distance,
 we consider revolution of the D3-branes around the fixed point. 
Then the gauge symmetry on D3-branes is spontaneously broken,
 and the rank of the gauge group is reduced. 
The distance can be set at our will
 by appropriately choosing the angular momentum of the revolving D3-branes,
 which should be determined by the initial condition of the cosmological evolution of D-brane configurations. 
The distance corresponds to the vacuum expectation values of brane moduli fields in the world-volume theory and,
 if it is written as $M/M_s^2$ in terms of the string scale $M_s$,
 the scale of gauge symmetry breaking is given by $M$.
Angular momentum conservation of revolving D3-branes assures the stability of the scale $M$ against $M_s$.
\end{abstract}

\end{center}
\end{titlepage}

\section{Introduction}
\label{sec:introduction}

The dynamics of electroweak symmetry breaking has become even more mysterious
 after the discovery of the Higgs boson.
It is widely believed that
 the standard model is merely 
 an effective theory of the electroweak symmetry breaking
 and some unknown physics or dynamics should exist behind it.
Many possibilities, mainly motivated by naturalness, have been examined, for example,
 dynamical symmetry breaking by strong coupling gauge interactions
 \cite{Weinberg:1979bn,Susskind:1978ms},
 radiative symmetry breaking with \cite{Ibanez:1982fr,Inoue:1982pi}
 or without supersymmetric extensions
 \cite{Coleman:1973jx,Meissner:2006zh,Foot:2007as,Iso:2009ss,Iso:2009nw,Iso:2012jn},
 Hosotani mechanism in the gauge-Higgs unification with extra dimensions
 \cite{Hosotani:1983xw}
 and so on. 

It is also an important issue to pursue 
 the origin of the masses of quarks and leptons,
 namely, how the electroweak symmetry breaking is mediated to them.
In the standard model it is simply described by  Yukawa couplings,
 and much efforts have been made to derive realistic Yukawa couplings.
Model buildings with D-branes in String Theory are most attractive,
 because Yukawa couplings can be understood by configurations of D-branes
 in the six-dimensional compact space of ten-dimensional superstring theories.
Some examples are intersecting D-brane models
 \cite{Cremades:2003qj,Cvetic:2003ch,Abel:2003vv,Kitazawa:2004hz,Kitazawa:2004nf,Blumenhagen:2006xt}
 or models of D-branes at singularities
 \cite{Aldazabal:2000sa,Berenstein:2001nk,Verlinde:2005jr,Krippendorf:2010hj}.
In these models, 
massless states of string with appropriate quantum numbers
 are identified with the Higgs doublet fields,
 but it will be generically difficult to obtain small vacuum expectation values 
 of the weak scale in the string theory setup.

In this paper we propose a new mechanism of spontaneous gauge symmetry breaking
 by assuming that our 3 dimensional space consists of revolving D-branes
 around a fixed point in 9+1 dimensional space-time.
As a simple model, we study D-branes at fixed points
 of a $T^6/{\bf Z}_3$ orbifold in type IIB superstring theory \cite{Aldazabal:2000sa}.
We assume that all the moduli, except for the D-brane moduli, are stabilized.
The six-dimensional torus is assumed to be factorizable, $T^6 = T^2 \times T^2 \times T^2$,
 and the action of ${\bf Z}_3$ is described by the twist vector $v = (1/3,1/3,-2/3)$.
There are three fixed points in each $T^2$ and  27 fixed points in total.
We distribute D3-branes and D7-branes and their anti-branes so as to
cancel the twisted R-R tadpoles  at each fixed point as well as
the untwisted R-R tadpoles  in the compact space.
As a result all supersymmetries in the type IIB superstring theory 
 are  broken through the ``brane supersymmetry breaking'' mechanism
 \cite{Sugimoto:1999tx,Antoniadis:1999xk,Angelantonj:1999jh,Aldazabal:1999jr,Angelantonj:1999ms}.
In particular, we consider a system of four D3-branes and three anti-D7-branes
 discussed in \cite{Kitazawa:2012hr}.
Three of four D3-branes can move away from the fixed point in a ${\bf Z}_3$ invariant way,
 and the separation of these D3-branes causes spontaneous gauge symmetry breaking of
 U$(2) \times $U$(1) \times $U$(1) \rightarrow$ U$(1) \times$U$(1)$
 on the world-volume theory on the D3-branes.
The rank of the gauge group is reduced
due to the identification of three D3-branes by the ${\bf Z}_3$ action.
The reduction of rank is actually necessary in the electroweak symmetry breaking.

When we put  anti-D7-branes on the fixed point as well as the D3-branes, 
there appears an attractive force between the D3-branes and the anti-D7-branes, and
when D3-branes move away from the fixed point, they are forced to be pulled back.
Thus  D3-branes tend to be localized at the fixed point, and 
the gauge symmetry breaking does not occur unless there exists an additional
balancing force that repels D3-branes from the fixed point.
In this paper, we consider revolution of the D3-branes whose centrifugal force balances
against the attractive force. 

In section \ref{sec:D-branes_at_singularities}
 we give a brief review of the models with D3-branes and D7-branes
 at the orbifold fixed points of a $T^6/{\bf Z}_3$ orbifold.
We explain
 how the vacuum expectation values of the D-brane moduli fields
 describe the displacements of the D3-branes. 
In section \ref{sec:rotating_D3-branes}
 the world-volume theory of revolving D3-branes is introduced.
The effect of revolution is an introduction of the centrifugal potential in 
the rest frame of the revolving D3-branes.
In section \ref{sec:symmetry_breaking}
 the spectrum in the world-volume theory is discussed.
We find that the spontaneous gauge symmetry breaking takes place and  
 the scale of the expectation value is stable against the string scale 
perturbations. In section \ref{sec:conclusion}, we conclude and discuss some 
important issues not studied in details in the present paper.

\section{D3 and D7-branes at $T^6/{\bf Z}_3$ orbifold fixed points}
\label{sec:D-branes_at_singularities}
\subsection{D-brane configurations}
General ideas and formulations of D-branes at orbifold fixed points
 are discussed in \cite{Aldazabal:2000sa}.
In this section
 we particularly consider the models with D3-branes, D7-branes and their anti-branes
 in a $T^6/{\bf Z_3}$ orbifold.
The six-dimensional torus is assumed to be factorized into three two-tori,
 $T^6 = T^2 \times T^2 \times T^2$ and 
 the complex coordinates of the corresponding two-tori are denoted by 
\begin{equation}
 Z_i = {1 \over \sqrt{2}} \left( X^{2i+2} + i X^{2i+3} \right)
 \label{defZi}
\end{equation}
 with $i=1,2,3$.
 Here $X^M$ with $M=0,1,\cdots,9$ are the coordinates of ten-dimensional space-time
 of type IIB superstring theory.
The coordinates are identified by the translations
\begin{equation}
 Z_i \sim Z_i + 2 \pi R_i,
\qquad
 Z_i \sim Z_i + 2 \pi R_i \tau,
\label{torus-identification}
\end{equation}
 where $R_i$ are the radii of the two-tori and $\tau = (-1 + i \sqrt{3})/2$.
We also identify points on the torus transformed by the ${\bf Z}_3$ action
\begin{equation}
 Z_i \sim e^{2 \pi i v_i} Z_i \ 
\label{Z3-identification}
\end{equation}
where $v = (1/3,1/3,-2/3)$ is the twist vector.
There are three fixed points in each two-torus under the above identifications,
 and  in total  27 fixed points  in the $T^6/{\bf Z}_3$ orbifold.
Since the string world-sheet fields
 should follow the same identifications as the coordinates $X^M$,
 there is an untwisted closed string propagating in the ten-dimensional space-time,
 and twisted closed strings localized at each fixed point.

We introduce D3-branes whose world-volume coordinates 
coincide with $X^\mu$ with $\mu=0,1,2,3$.
In six-dimensional compact space
 the positions of these D3-branes are specified by the corresponding points
 which are distributed in a ${\bf Z}_3$ symmetric way in the  torus.
Therefore, when D3-branes are put away from the fixed points,
three D3-branes must move together which belong to the following
 regular representation of ${\bf Z}_3$.
The cyclic permutation of three D3-branes in the  torus is described by 
the matrix
\begin{equation}
 \gamma = \left(
           \begin{array}{ccc}
            0 & 1 & 0 \\
            0 & 0 & 1 \\
            1 & 0 & 0
           \end{array}
          \right),
\end{equation}
which can be diagonalized by a unitary transformation as
\begin{equation}
 U \gamma U^\dag = \left(
                    \begin{array}{ccc}
                     1 & 0 & 0 \\
                     0 & \alpha & 0 \\
                     0 & 0 & \alpha^2
                    \end{array}
                   \right)
\label{regular}
\end{equation}
 where $\alpha = \exp(2 \pi i /3)$.
This shows that the regular representation is not an irreducible representation of ${\bf Z}_3$.
A D3-brane in the irreducible representation of ${\bf Z}_3$
with one of the above three eigenvalues of $\gamma$ is called a fractional D3-brane.
It is localized at the fixed point and cannot move away from it.
Only a set of three D3-branes can move. 
The above matrices also act as the ${\bf Z}_3$ operations on the Chan-Paton indices of the open strings
 whose ends are confined in the corresponding D3-brane world-volume.

The twisted R-R tadpole cancellation condition of the D-branes at a fixed point
is given by using the matrices of the ${\bf Z}_3$ operation acting on 
 the various D-branes:
\begin{equation}
 3 \, {\rm Tr} (\gamma_3) - 3 \, {\rm Tr} (\gamma_{\bar 3}) + {\rm Tr} (\gamma_7) - {\rm Tr} (\gamma_{\bar 7}) = 0.
 \label{RRcan}
\end{equation}
Here, $\gamma_3, \gamma_{\bar 3}, \gamma_7$ and $\gamma_{\bar 7}$ are the matrices
for D3, anti-D3, D7 and anti-D7 branes localized at the fixed point.
A simple solutions to the condition of eq.~(\ref{RRcan}) with four D3-branes and three anti-D7-branes
 is given by
\begin{equation}
 \gamma_3 = {\rm diag} ({\bf 1}_2, \alpha, \alpha^2),
\qquad
 \gamma_{\bar 7} = {\bf 1}_3,
 \label{fourD3AD7}
\end{equation}
 where ${\bf 1}_N$ denotes an $N \times N$ unit matrix \cite{Kitazawa:2012hr}.
Among four D3-branes at the fixed point,  
 a set of three can move away from the fixed point by constructing 
 the regular representation of eq.~(\ref{regular}).
The remaining D3 is in the 
 irreducible representation ${\bf 1}_1$ with an eigenvalue $1$ of the matrix $\gamma$.

\begin{figure}[t]
\centering
\includegraphics[width=140mm]{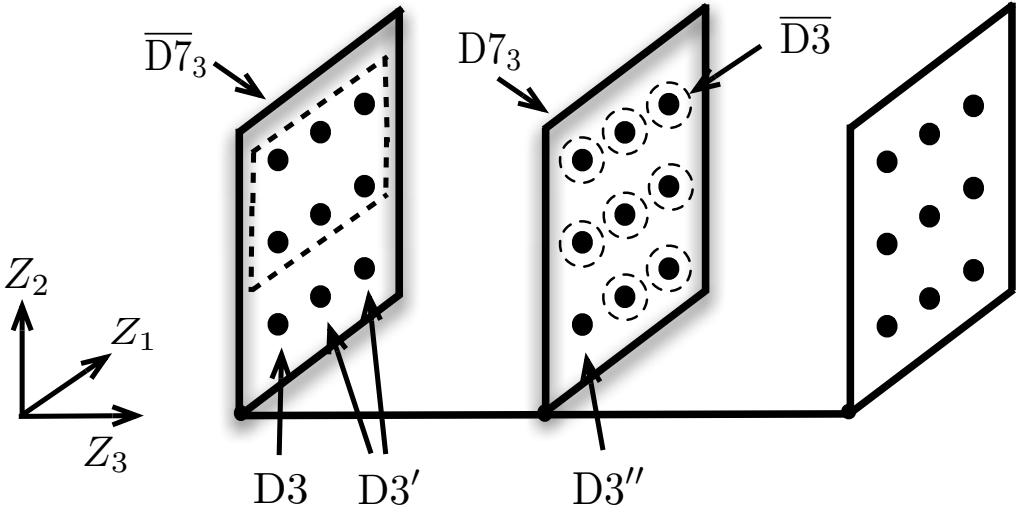}
\caption{
An example of D-brane models in which all the twisted and untwisted R-R tadpoles are cancelled.
Each of the complex coordinates of three tori $Z_i$
 is represented as a line and blobs on each line indicates three fixed points on the torus.
D7 and anti-D7 branes are not extended to the third torus.
Owing to the Wilson line introduced on the anti-D7-branes,
no D3-branes or anti-D3-branes are required
 for six fixed points inside the square with a dashed line
 (see text).
}
\label{fig:config}
\end{figure}

The above subsystem with eq.~(\ref{fourD3AD7}) 
can be embedded globally in the $T^6/{\bf Z}_3$ orbifold
as shown in Figure~\ref{fig:config}.
Since the codimension of D7-branes is 2 in ten-dimensional space-time,
we can specify it by indicating directions along which
its world-volume is not extended. 
D$7_i$ denotes  a D7-brane whose world-volume does not include 
the direction of $i$-th torus described by  $Z_i$. 
The ${\bf Z}_3$ action matrices for other D-branes are given by
\begin{equation}
 \gamma_{3'} = {\bf 1}_1,
\qquad
 \gamma_{3''} = {\rm diag} ( {\bf 1}_1, \alpha ),
\qquad
 \gamma_{\bar 3} = \alpha^2,
\qquad
 \gamma_7 = \alpha^2 {\bf 1}_3.
\end{equation}
All of the above D-branes are fractional D-branes.
We also introduce Wilson line on the anti-D7-branes 
 which is described by the operation matrix
\begin{equation}
 \gamma_W = {\rm diag} ( 1, \alpha, \alpha^2),
\end{equation}
 corresponding to the torus shifts of eq.~(\ref{torus-identification}).
The twisted R-R tadpole cancellation conditions for the fixed points,
 which are fixed under ${\bf Z}_3$ transformations up to the shift of eq.~(\ref{torus-identification}),
 are modified by replacing $\gamma_{\bar 7}$ by $\gamma_W\gamma_{\bar 7}$ or $\gamma_W^2\gamma_{\bar 7}$
 in eq.~(\ref{RRcan}) \cite{Aldazabal:1998mr}.

The untwisted R-R tadpoles are cancelled out,
 because the numbers of D3-branes and anti-D3-branes are the same,
 and the numbers of D7-branes and anti-D7-branes are the same.
If moduli stabilization requires some other 
objects which are sources of untwisted R-R charges
 \cite{Kachru:2003aw,Kachru:2003sx,Antoniadis:2004pp,Antoniadis:2006eu,Kitazawa:2014hya},
 this D-brane configurations are changed accordingly.
In the following analysis, 
 we simply assume that the model is constructed in a globally consistent way
 and focus on the subsystem of eq.~(\ref{fourD3AD7})
near the fixed point where four D3 and three $\overline{{\rm D}7}_3$ branes are localized.
Since the supersymmetry is broken in the present model,
 the NS-NS tadpoles are not cancelled
 and the flat space-time with trivial dilaton and B-field background
 is no longer a solution to the equation of motion of 
 string theory \cite{Fischler:1986ci,Fischler:1986tb}.
The effects of the NS-NS tadpoles on the background geometry
 are neglected for the moment and will be mentioned briefly
 in section \ref{sec:conclusion}.

\subsection{DBI action}
A bosonic part of the world-volume low-energy effective theory
 of a D$p$-brane is given by the generalized DBI action 
\begin{equation}
 S_p = - \tau_p \int d^{p+1}\xi \, e^{-\phi}
         \sqrt{-{\rm det} \left( G_{ab} + B_{ab} + 2 \pi \alpha' F_{ab} \right)},
\label{DBI}
\end{equation}
 where $\tau_p = 1/(2\pi)^p g_s (\alpha')^{(p+1)/2}$ is the tension of the D$p$-brane
 while string coupling $g_s = e^{\langle \Phi \rangle}$
 is determined by the vacuum expectation value of the ten-dimensional dilaton field
 $\Phi = \langle \Phi \rangle + \phi$.
The induced metric on the D$p$-brane is defined as
\begin{equation}
 G_{ab} = {{\partial X^M} \over {\partial \xi^a}} {{\partial X^N} \over {\partial \xi^b}} \eta_{MN},
\end{equation}
 where $X^M(\xi)$ describe the embedding of 
 the D$p$-brane in ten-dimensional space-time.
Here, we assume that  the ten-dimensional  space-time is flat Minkowski.
Generalizations of the DBI action to $n$ D$p$-branes have been  discussed in various literatures.

Let us now focus on the  $p=3$ case.
 The low-energy effective action of $n$ D3-branes is described by
 four-dimensional ${\cal N}=4$ U$(n)$ super-Yang-Mills theory
 whose bosonic part is obtained by expanding the non-Abelian 
 generalization of the DBI action in eq.~(\ref{DBI}).
The vector multiplet of four-dimensional ${\cal N}=4$ U$(n)$ super-Yang-Mills theory
 consists of U$(n)$ gauge bosons, four Weyl fermions and six real scalar fields.
The six scalar fields are originated from the embedding fields $X^M(\xi)$ 
with $M=4,5, \cdots , 9$
and can be interpreted as the brane moduli fields 
 whose vacuum expectation values 
 describe the positions of D-branes in six-dimensional compact space.
 We define the complex combinations $Z_i$ ($i=1,2,3$) of these
 moduli fields as in eq.~(\ref{defZi}).
 In the absence of other sources of supersymmetry breaking such as the
anti-D7-branes, there is no potential along the moduli directions.
As we will see, 
 some of these properties survive after the $T^6/{\bf Z}_3$ compactification,
 namely in the world-volume theory of D3-branes at the orbifold fixed points.

\subsection{Gauge symmetry breaking in the quiver gauge theory}
\begin{figure}[t]
\centering
\includegraphics[width=50mm]{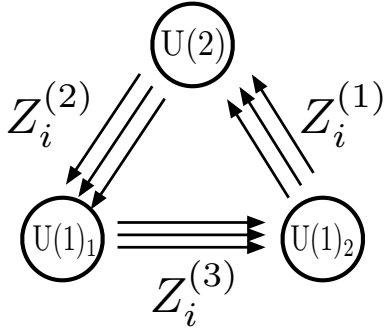}
\caption{
The world-sheet theory of the four D-branes at the orbifold fixed point
 is described by the U$(2)\times $U$(1)_1 \times $U$(1)_2$ quiver gauge theory. 
The fields $Z_i^{(a)}$ are in the bi-fundamental representations
 and depicted by arrows between the blobs.
Three lines with $i=1,2,3$ correspond to complex fields on each of the three tori.  
}
\label{fig:quiver}
\end{figure}

The world-volume theory of the four D3-branes at the orbifold fixed point
 can be obtained by imposing the orbifolding conditions on ${\cal N}=4$ SYM theory
 and described by the four-dimensional ${\cal N}=1$ U$(2) \times$U$(1)_1 \times$U$(1)_2$ 
 quiver gauge theory as shown in Figure~\ref{fig:quiver}.
It has the chiral multiplets in the bi-fundamental representations
\begin{center}
\begin{tabular}{cccc}
 & U$(2)$ & U$(1)_1$ & U$(1)_2$ \\
 $Z^{(1)}_i$ & $2$ & $0$ & $-1$ \\
 $Z^{(2)}_i$ & $2^*$ & $+1$ & $0$ \\
 $Z^{(3)}_i$ & $1$ & $-1$ & $+1$ \\
\end{tabular}  
\end{center}
The subscript $i$ with $i=1,2,3$ denotes the index of three tori $Z_i$
 and the superscript $(a)$ with $a=1,2,3$ denotes three different types of
 bi-fundamental representations of the U$(2) \times$U$(1)_1 \times$U$(1)_2$ gauge symmetry.
We use the same symbols $Z_i^{(a)}$ to describe the scalar components of the corresponding chiral superfields.
The covariant derivatives on the scalar fields are given by
\begin{eqnarray}
 D_\mu Z_i^{(1)} &=&
  \left(
   \partial_\mu Z_i^{(1)}
   + i g \left(\frac{\tau^A}{2} \right) W_\mu^A Z_i^{(1)}
   - i\frac{g}{\sqrt{2}} Z_i^{(1)} B_\mu^{(2)}
  \right)
\n
 D_\mu Z_i^{(2)} &=&
  \left(
   \partial_\mu Z_i^{(2)}
   + i g \left(\frac{-\tau^A}{2} \right)^* W_\mu^A Z_i^{(2)}
   + i\frac{g}{\sqrt{2}} Z_i^{(2)} B_\mu^{(1)}
  \right)
\n
 D_\mu Z_i^{(3)} &=&
  \left(
   \partial_\mu
   - i\frac{g}{\sqrt{2}} B_\mu^{(1)} 
   + i\frac{g}{\sqrt{2}} B_\mu^{(2)} \right) Z_i^{(3)},
\label{Cov-Der}
\end{eqnarray}
 where the gauge fields of U$(2)$, U$(1)_1$ and U$(1)_2$ are described as 
 $W^{A}_\mu, B^{(1)}_\mu$ and $B^{(2)}_\mu$ with $A = 0,1,2,3$.
 The $\tau$ matrices are defined as $\tau^A=( {\bf 1}_2, \sigma^i)$ where
 $\sigma^i$ are the Pauli matrices.
The gauge coupling constant is defined as $g = \sqrt{2 \pi g_s}$.
Among the U$(1)$ symmetries,
 the combination of U$(1)_1 - $U$(1)_2$ is the anomalous U$(1)$ symmetry
 whose chiral anomaly is cancelled by the 
 generalized Green-Schwarz mechanism \cite{Ibanez:1998qp}.
 
These scalar fields acquire the F-term potential
\begin{equation}
 V_F = \sum_{i,a} \left|  \frac{\partial W}{\partial Z_i^{(a)} }\right|^2,
\label{F-potential}
\end{equation}
 where $W$ is given by
\begin{equation}
 W = i g \epsilon_{ijk} Z^{(2) \beta}_i Z^{(1)}_{j\beta} Z^{(3)}_k,
\label{superpotential}
\end{equation}
 and $\beta$ is the index of the doublet representation of the U$(2)$ group. 
There are also the D-term potential
\begin{equation}
 V_D = {1 \over 2} \left( (D_{{\rm U}(2)}^a )^2 + (D_1)^2 + (D_2)^2 \right),
\end{equation}
 where 
 \begin{eqnarray}
 D_{{\rm U}(2)}^A &=&
  - g \left(
       Z^{(1)\dag}_i {{\tau^A} \over 2} Z^{(1)}_i
       + Z^{(2)\dag}_i \left( -{{\tau^A} \over 2} \right)^* Z^{(2)}_i
      \right), \\
 D_1 &=&
  - g \left(
       Z^{(2)\dag}_i {1 \over \sqrt{2}} Z^{(2)}_i
       + Z^{(3)\dag}_i \left( -{1 \over \sqrt{2}} \right) Z^{(3)}_i
      \right), \\
 D_2 &=&
  - g \left(
       Z^{(3)\dag}_i {1 \over \sqrt{2}} Z^{(3)}_i
       + Z^{(1)\dag}_i \left( -{1 \over \sqrt{2}} \right) Z^{(1)}_i
      \right).
\end{eqnarray}
The normalization of the Lie generators is fixed by the relation
${\rm Tr} (T^a T^b) = \delta^{ab} /2$.
The flat directions of the potential $V_F+V_D$
 are parametrized by the six parameters $(v_i, \theta_i)$ as
\begin{equation}
 \langle Z^{(1)}_i \rangle =
  \left( \begin{array}{c} 0 \\ v_i e^{i \theta_i} \end{array} \right),
\qquad
 \langle Z^{(2)}_i \rangle =
  \left( \begin{array}{c} 0 \\ v_i e^{i \theta_i} \end{array} \right),
\qquad
 \langle Z^{(3)}_i \rangle = v_i e^{i \theta_i}.
\label{flat-general}
\end{equation}
The relative phases between $Z_i^{(a)}$ and $Z_i^{(b)}$ with different $a$ and $b$
 can be absorbed by some combinations of gauge transformations
 of U$(1)$ component of U$(2)$ and U$(1)_1$ and U$(1)_2$,
 though the common phase cannot be gauged away,
 because all the fields are in bi-fundamental representations.
If the set of three D-branes gets the vacuum expectation value of eq.~(\ref{flat-general}),
 the gauge symmetry U$(2) \times $U$(1)_1 \times$U$(1)_2$ is spontaneously broken  down to U$(1) \times $U$(1)$.
Expanding around the vacuum expectation value,
 the gauge bosons associated with the broken generators acquire the mass of $\sqrt{\sum_i g^2 v_i^2}$.

\begin{figure}[t]
\centering
\includegraphics[width=50mm]{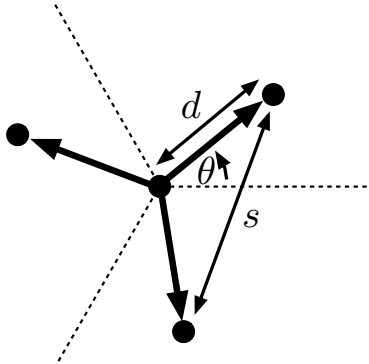}
\caption{
A ${\bf Z}_3$ symmetric separation of three D3-branes away from the fixed point in $T^2$.
Blobs describe D3-branes.
There are two distances between D3-branes: $d$ and $s=\sqrt{3}d$.
}
\label{fig:separation}
\end{figure}

It has the following geometrical interpretation. 
These vacuum expectation values
 describe the configuration of three D3-branes away from the fixed point on $T^6/{\rm\bf Z}_3$.
They are distributed in a ${\bf Z}_3$ invariant way on each torus
 as shown in Figure~\ref{fig:separation}.
Since each $Z_i^{(a)}$ is a scalar mode of an open string stretching from a D-brane to another,
 the mass is proportional to the distances, $d$ or $s$, namely,
 $d/2\sqrt{2}\pi\alpha'$ or $s/2\sqrt{2}\pi\alpha'$, respectively.
Writing the position of three D-branes on $i$-th torus as 
 $(d_i e^{i \theta_i}, d_i e^{i (\theta_i+2\pi/3)}, d_i e^{i (\theta_i+4\pi/3)})$,
 the relative complex coordinates between them are given by, e.g.,
 $d_i e^{i\theta_i} (e^{i 2\pi/3} -1) =  i \sqrt{3} d_i e^{i (\theta_i +\pi/3)}$.
If we take a different pair of D-branes, 
the constant part of the phase $\pi/3$ is changed. 
The distance $s$ between the branes is related to the distance of 
 the branes from the origin $d=\sqrt{\sum_i d_i^2}$ by the relation
 $s =\sqrt{3} d$ as shown in Figure~\ref{fig:separation}.
The geometric interpretation of the coordinates and boson masses
 is made more explicit in the following calculations, but
 we want to stress here that the total phase 
 $\theta_i$ represents the direction of the set of three D3-branes, as shown in 
 Figure~\ref{fig:separation}.　

Within the ${\cal N}=1$ quiver gauge theory, there is no potential along 
the direction of increasing $v_i$ and the vacuum expectation 
value cannot be dynamically determined.
In the next section, 
in order to study the dynamics of the gauge symmetry breaking in string theory, 
we will include the effect of the anti-D7-branes at the fixed point
 as shown in Figure~\ref{fig:config}, which breaks the supersymmetry.
Furthermore
 we introduce revolution of D3-branes to balance the attractive force between D3 and anti-D7 branes.
These two effects determine the vacuum expectation value $v_i$.

\section{Revolving D3-branes around anti-D7-branes}
\label{sec:rotating_D3-branes}

In this section
 we study the dynamics of the gauge symmetry breaking 
 in the model of revolving D3-branes around the anti-D7-branes put at the orbifold fixed point.
It is realized  as a subsystem in the configuration of Figure~\ref{fig:config}.
\subsection{Attractive potential between ${\rm D}3$ and $\overline{{\rm D}7}$ branes}
First we consider an effect of the anti-D7-branes.
This breaks the supersymmetry of the D3-brane system.
Since the anti-D7-branes are not extended along the third torus $Z_3$, 
 it can be expected that the D3-brane flat direction in eq.~(\ref{flat-general})
 is lifted by an appearance of a mass-like term along $Z_3^{(a)}$. 
The supersymmetry breaking potential cannot be understood by the closed string exchange,
 since the situation we have in mind is $d<l_s$:
 the distance $d$ between the D3-branes and the anti-D7-branes
 is much shorter than the string length $l_s \equiv \sqrt{\alpha'}$ (see ref. \cite{Dvali:1998pa}).
Instead,
 we can estimate it by calculating one-loop corrections of open strings
 stretching between D3 and anti-D7 branes, whose spectra do not preserve supersymmetry.
For $|Z_3^{(a)}| \ll M_s$,
 the supersymmetry breaking potential can be expanded and the mass term
\begin{equation}
 \mu^2 \sum_a |Z_3^{(a)}|^2
 \label{masspotential}
\end{equation} 
 is added to the F-term and D-term potentials $V_F+V_D$. 
An order estimation \cite{Kitazawa:2012hr} gives the coefficient
\begin{equation}
 \mu^2 = {1 \over {C^2}} {{g^2} \over {16 \pi^2}} M_s^2,
\end{equation}
 where $M_s \equiv 1/\sqrt{\alpha'}$ and $C$ is a constant $C \sim {\cal O}(1)$. 
 
Since D3-branes are attracted to the anti-D7-branes due to the potential (\ref{masspotential}), 
 they are bound to the anti-D7-branes at the fixed point on the third torus
 unless another repulsive interaction is added. 
Other anti-D-branes can be introduced far from the fixed point
 so that D3-branes are separated apart from the fixed point. 
But, it is generally difficult to stabilize the positions of D3-branes,
 and furthermore, even if they are stabilized,
 the typical length scale is given by the string scale,
 since there are no other dimensionful parameters in the theory. 

\subsection{A toy model of revolution}
\label{sec:toy}
In the following,
 we consider D3-branes revolving around an anti-D7-branes 
 whose centrifugal force balances the attractive force from the anti-D7-branes.
As a warm up exercise,
 let us consider a single complex scalar field $\phi$
 which represents the moduli of the set of three D3-branes on the third torus,
 namely, each component of $Z^{(a)}_3$ which has the vacuum expectation value
 (see eq.~(\ref{flat-general})).
The Lagrangian we consider is simply given by
\begin{equation}
 {\cal L} = |\dot\phi|^2-|\nabla \phi|^2 - \mu^2 |\phi|^2 .
\label{massive-scalar}
\end{equation}
where the mass term comes from the attraction between the D3 and anti-D7 branes. 
The coefficient is $\mu \sim M_s/10$.
When we study time-dependent solutions,
 it is more convenient to consider the Hamiltonian density,
\begin{equation}
 {\cal H} =|\pi_\phi|^2+|\nabla \phi|^2 + \mu^2 |\phi|^2 .
\label{H-massive-scalar}
\end{equation}
Here, $\pi_\phi =\dot \phi^\dagger$.
A solution to the equation of motion is given by 
\begin{equation}
 \phi = v_0 \exp(i \omega t)
\label{classical-solution-massivescalar}
\end{equation}
 with $\omega=\mu$.
Since $\phi=(\phi_1 + i \phi_2)/\sqrt{2}$ represents the position of three D3-branes,
 this solution can be interpreted as a revolving D3-brane with an angular velocity $\omega$.
In quantizing the system,
 we usually treat the  zero mode wave function equally with the other plane waves with nonzero momenta,
 because we consider that $\phi$ itself represents a microscopic quantum field. 
The amplitude of the zero-mode $v_0$ fluctuates around $v_0=0$ and 
takes various values.
But what if the solution is macroscopic (or classical)? 
In our case, $\phi$ represents the position of a D3-brane
 and the zero mode  wave function $\phi = v_0 \exp(i \omega t)$ has a geometrical meaning 
 of the revolving D3-brane around the center;
 $(\phi_1, \phi_2)=(\sqrt{2} v_0 \cos \omega t, \sqrt{2} v_0 \sin \omega t)$.
Then the solution $\phi$ must be treated classically as the coordinate of the revolving D3-brane.
We thus need to expand the scalar field around the solution 
 (\ref{classical-solution-massivescalar}) and quantize the fluctuations,  
 just as we do in the case of the Higgs field around a nontrivial vacuum in a double well potential. 

The next question is what determines the value of $v_0$.
It can not be fixed by the equation of motion derived from the Lagrangian (\ref{massive-scalar}).
The  situation is different from the case of the ordinary Higgs 
 where it acquires a nontrivial vacuum expectation value
 at the  minimum of the effective potential\footnote{In the case
 of the ghost condensation \cite{ArkaniHamed:2003uy}, the derivative of a scalar field
 acquires a nonzero value as a result of the equation of motion.
 See discussions in section \ref{sec:conclusion}
 for comparison between the present model and the ghost condensation.}.
In the present case, the would-be moduli field will acquire a nonzero value of $v_0$
 as a result of the angular momentum conservation of the revolving D3-brane.
In this sense,
 the determination of $v_0$ needs some external input (or an initial condition)
 provided from outside of the four dimensional field theory. 
In order to see it, we first note that 
 the Lagrangian (\ref{massive-scalar}) is invariant under the global U$(1)$ phase
 rotation $\phi \rightarrow e^{i \theta} \phi$.
The corresponding Noether current 
\begin{equation}
 j_\mu = -i \left( \phi^* \partial_\mu \phi - \partial_\mu \phi^* \phi \right)
\end{equation}
 is conserved; $\nabla^\mu j_\mu=0$. 
In the geometrical interpretation in terms of the D3-brane configurations, 
 the symmetry is nothing but the rotational symmetry of the D3-brane around the center.
Inserting $\phi = v_0 \exp(i \omega t)$ into the current conservation,
 we get the conservation of the angular momentum of the revolving D3-brane around the anti-D7-branes;
\begin{equation}
 {\partial \over {\partial t}} \left( v_0^2 \omega \right) = 0.
 \label{AMconservation}
\end{equation}
The quantity $l=v_0^2 \omega$ is a constant of motion
with a mass dimension 3,
 and related to the angular momentum of the revolving D3-brane.
The angular momentum is given by $M_{\rm D3} \omega d^2$,
 where $M_{\rm D3} \equiv \tau_3 V_{\rm D3}$ is the mass of D3-brane
 ($V_{\rm D3}$ is the volume of the D3-brane)
 and $d$ is the distance from the center of the revolution.
The mass $M_{\rm D3}$ can be read from the kinetic term in the DBI action of eq.~(\ref{DBI}),
 where the scalar fields $X^M$ are regarded as the coordinates of the brane embedded 
 in the ten-dimensional space-time.
The relations between $d$ and $v_0$, $d = 2\sqrt{2} \pi \alpha' g v_0$,
 are obtained by identifying the mass of the ground state of stretched open string, $d/2\sqrt{2} \pi \alpha'$,
 and gauge boson mass through the Higgs mechanism, $g v_0$.
Hence, the value of $l$, which determines the value of $v_0$, 
needs to be fixed by the initial condition
 of the cosmological evolution of D3-brane configurations.
At the end of the discussions in section~\ref{sec:conclusion},
 we generalize the analysis 
 to include the scale factor $a(t)$ of the FRW universe on D3-branes.
The essential points discussed here are not changed.
  
In order to see the appearance of the centrifugal potential,
 we focus on spacial homogeneous solutions and write the complex scalar field as $\phi = v(t) e^{iu(t)}$.
It can be expanded as $\phi= (v_0+\delta v) \exp[ i (\omega t+\pi) ]=v_0 e^{i \omega t} + \varphi$,
 where $\varphi \sim (\delta v + i v_0 \pi) e^{i \omega t}$.
By inserting $\phi$ into the Hamiltonian density, we get
\begin{eqnarray}
 {\cal H}
  &=&  |\dot{v}+iv \dot{u}|^2 + \mu^2 v^2
\n
  &=& (\delta \dot{v})^2 + (v_0+\delta v)^2 (\omega+\dot{\pi})^2 + \mu^2 (v_0+\delta v)^2.
\label{H-toy-model}
\end{eqnarray}
On the other hand, the current conservation requires that the combination
\begin{equation}
 l =  v^2 \dot{u} =  (v_0+\delta v)^2 (\omega + \dot{\pi}) 
 \label{l-conservation}
\end{equation}
 is a constant of motion which should be fixed by an initial condition. 
This introduces a constraint in the system, and reduces the dynamical degrees of freedom.
The Hamiltonian is written in terms of $\delta v$ only and becomes
\begin{equation}
 {\cal H}  =  (\delta\dot{v})^2 + \frac{l^2}{ (v_0+\delta v)^2} + \mu^2 (v_0+\delta v)^2.
\label{revolvingHam}
\end{equation}
Due to the conservation of the angular momentum, the phase fluctuation $\pi$
disappears from the Hamiltonian. It is physically reasonable since the 
conservation of angular momentum relates the angular velocity
with the radius of the revolution. Of course, this reduction of dynamical degrees of 
freedom is applied only to the zero mode, and non-zero momentum modes 
of $\pi$ do not 
disappear from the Hamiltonian because only the total angular momentum is 
conserved. We will see this explicitly in section \ref{sec:symmetry_breaking}.
Hence the role of the angular momentum conservation is 
not the reduction of degrees of freedom, but the determination of the value of $v_0$.
The minimum of the potential density
$V=l^2/ v_0^2+ \mu^2 v_0^2$ determines 
the value of $v_0$ as $ v_0^2=l/\mu$. 
It is consistent with the relation 
$l= v_0^2 \omega$ and
the frequency of the classical solution of the revolution $\omega=\mu$.

In the  toy model discussed above, we considered only the zero mode of field configuration
which does not depend on the space coordinates. 
In obtaining the centrifugal potential for nonzero modes, 
we need to care about  competition between the
strength of the centrifugal force and the tension on the brane.
There are two limiting cases depending on which force is stronger than the other.
If the centrifugal force is stronger than the tension, each point on 
the brane behaves as if it is revolving independently of the other points.
In this case, we can use an approximation that
the conservation of the
angular momentum at each point  holds separately. Then each point
is bound to the minimum of the potential in  (\ref{revolvingHam}).
The coefficient $l^2$ of the centrifugal potential can depend on 
positions $x$, and hence so is the minimum $v_0(x)$.
Due to the balance between the
attractive force $F_{\rm attractive} =\mu v(x)$ and the repulsive force 
$F_{\rm repulsive}=\omega(x)  v(x)$ at each point on the D3-brane,
the brane revolves with the common angular frequency $\omega(x)=\mu$
independent of the value of $v_0(x)$. 
On the other hand, in the opposite case where the tension is stronger,
the only conserved quantity is the total angular momentum,
and  integral of (\ref{l-conservation}) over the space,
$L=\mu T_{\rm D3} \int v(x)^2 d^3 x.$
Then the brane can rotate around itself as it revolves.
The degrees of freedom on the D3-brane can contribute to the angular momentum.
In the following, we consider the first situation.  Then each point
on the brane is strongly constrained in the minimum of the potential. 
Furthermore, we consider the simplest case where $v_0(x)$ is independent of $x$
\footnote{If the attractive potential is deviated from the harmonic potential, the angular
frequency $\omega(x)$ depends on $v_0(x)$. A single brane favours to revolve
with the same angular velocity in order to avoid ripping. Hence $v_0(x)$ may have
tendency to be aligned.}.
General situations are left for future investigations.

\subsection{Revolving D3-brane in DBI action}
When the embedding $X^\mu$ represents revolution of a D3-brane, we can see that
 the DBI action (\ref{DBI}) is reduced to the toy model discussed above.
In the static gauge, 
 a revolving D3-brane on the third torus is described by the
 following set of coordinates;
\begin{equation}
 X^\mu = \xi^\mu,
\qquad
 Z_3= (\tilde{d} e^{i \omega \xi^0} + \varphi ) /\sqrt{\tau_3} 
\end{equation}
where  ${\tilde d} \equiv \sqrt{\tau_3} d$. 
We set $Z_1=Z_2 = 0$ for simplicity
since we are interested in revolution of the D3-brane on the third torus. 
Then the induced metric on the world-volume of the D3-brane is given by
\begin{equation}
 G_{ab} = \eta_{ab}
  + {1 \over {\tau_3}} \tilde{D}_a\varphi^\dag  \tilde{D}_b\varphi
\end{equation}
 where
\begin{equation}
 \tilde{D}_0\varphi = {{\partial \varphi} \over {\partial \xi^0}} + i {\tilde d} \omega e^{i\omega\xi^0},
\qquad
 \tilde{D}_i\varphi = {{\partial \varphi} \over {\partial \xi^i}}.
\end{equation}
Setting B-field, dilaton and gauge fields zero, the DBI Lagrangian becomes
\begin{eqnarray}
 {\cal L}_{\rm DBI}
 &=& -\tau_3 \sqrt{-{\rm det} G_{ab}}
\nonumber\\
 &=& -\tau_3 
 + |\tilde{D}_0 \varphi|^2 -|\nabla \varphi|^2 + {\cal O}(\varphi^4)
\label{L-DBI-revolution}
\end{eqnarray}
Including the supersymmetry breaking mass term (\ref{masspotential}), 
 this is the same as the toy model of the massive free scalar field of eq.~(\ref{H-toy-model})
 by identifying the parameter as $ {\tilde d} = \sqrt{2} v_0$.
Then we obtain the relation $d = 2 \sqrt{2} \pi \alpha' g v_0$. 
A generalization to $n$ D3-branes is straightforward. 

\subsection{Effects of $\overline{{\rm D}7}$-branes and revolution of D3-branes in $T^6/{\bf Z}_3$ }
The effects of 
  the anti-D7-branes and the revolution of D3-branes in 
  the world-volume field theory of the D3-branes can be summarized by
adding a potential term  $l^2/ v_0^2 + \mu^2 v_0^2$ to the world-volume
field theory of D3-branes.
Here $v_0$ is proportional to the distance between the anti-D7 and D3 branes.

In the rotating reference frame where D3-branes are at rest, 
 the branes feel the centrifugal potential $l^2/| \phi|^2$. 
Here,  $\phi$ is the moduli field representing the complex coordinate of the D3-branes.
Originally it comes from the kinetic term
 but we can regard the term as one of the potential terms 
 in the rotating reference frame.
The discussion can be straightforwardly generalized
 to the case of revolving D3-branes on the $T^6/{\rm Z}_3$ orbifold. 
Since the three D3-branes are interacting each other, 
 they exchange angular momenta so that
 only a sum of each angular momentum of D3-branes on the third torus is conserved. 
The conserved current is then a sum of each angular momentum of D3-branes
\begin{equation}
 j_\mu = -i \sum_a \left( Z_3^{(a)*} \partial_\mu Z_3^{(a)} - 
 \partial_\mu  Z_3^{(a)*}  Z_3^{(a)} \right) .
\end{equation}
Following the same argument in the toy model, the effects of the anti-D7-branes 
and the revolution
of three D3-branes are to
generate the attractive and the repulsive potentials respectively;
\begin{equation}
V_M(Z) = \frac{(3l)^2}{ \sum_a |Z_3^{(a)}|^2} + 
\mu^2 \sum_a |Z_3^{(a)}|^2 .
\label{VM}
\end{equation}
The coefficient $3$ in front of $l$ is introduced for later convenience.
In order to apply this to nonzero modes, as we discussed at
the end of section \ref{sec:toy}, the centrifugal force must be stronger than
the tension.
In the next section, we discuss the low energy spectrum of the 
D3--anti-D7 brane system with the above potential.
Since the potential $V_M(Z)$ depends only on the combination 
$\sum_a |Z_3^{(a)}|^2$, mass term appears only in the {\it breathing} mode that
changes the magnitude of the same combination.
In the rotating reference frame on the third torus, 
without loss of generality, we can choose the classical solution of D3-branes
$Z_i^{(a)}$ as
\begin{equation}
 \langle Z^{(1)}_3 \rangle =
  \left( \begin{array}{c} 0 \\ v_0  \end{array} \right),
\qquad
 \langle Z^{(2)}_3 \rangle =
  \left( \begin{array}{c} 0 \\ v_0 \end{array} \right),
\qquad
 \langle Z^{(3)}_3 \rangle = v_0,
 \qquad
 \langle Z^{(a)}_1 \rangle =\langle Z^{(a)}_2 \rangle =0.
\label{Z-sol-revolv}
\end{equation}
Here, we assumed that
 there is no separations of D3-branes in the first and the second tori.
This satisfies the D-flat condition.
Note that
 we are now in the rotating reference frame of the D3-branes
 and the vacuum expectation value is set stationary without the phase. 
The breathing mode $\sigma$ is defined as that 
 to change the magnitude of $\sum_a |Z_3^{(a)}|^2$,
\begin{equation}
 v_0 \rightarrow v_0 + \frac{\sigma}{\sqrt{6}} .
\label{breathing-mode}
\end{equation}
The normalization of the $\sigma$ field is chosen so that the kinetic term becomes canonical.
Inserting this into the potential $V_M$,
 we can read the mass $m_\sigma$ of the breathing mode as $m_\sigma = 2\mu$.
 Hence it becomes as heavy as the string scale.
The other modes are massless. 

Finally in this section, let us play with the numerics of various quantities. 
D3-brane is a gigantically macroscopic object
 and its mass is obtained as $M_{\rm D3}=\tau_3 V_{\rm D3}$.
If we set $M_s=10^{18} \ {\rm GeV}$ and $V=(10^{-33} \ {\rm eV})^{-3}$
 (i.e. the present particle horizon),
 the mass becomes $M_{\rm D3}=10^{207} \ {\rm eV}$. 
Of course,
 it is much larger than the corresponding ``mass''
 obtained from the critical density $\rho_{\rm crit}$ of our current universe
 $\rho_{\rm crit} V_{\rm D3} \sim 10^{89} \ {\rm eV}$.
It is nothing but the cosmological constant problem,
 since the energy scale of the tension of D3-brane is typically as large as the string scale 
 and much larger than the energy scale of the dark energy at the present universe, i.e. meV.
The total angular momentum of the D3-brane is thus given by 
\begin{equation}
 L = M_{\rm D3} \omega d^2 \sim (gv_0)^2 M_s V_{\rm D3}.
\end{equation}
In the second equality,
 we dropped numerical factors and used  $\omega = \mu = M_s$ and $d^2 = (gv_0)^2 /M_s^4$.
When we put, e.g. $gv_0=100 \ {\rm GeV}$, the angular momentum is roughly $L \sim 10^{148}$.
Hence, we can treat the revolution of the D3-brane classically. 
Though the angular velocity $\omega$ is as huge as the string scale,
 the velocity of the revolution is non-relativistic
 if the vacuum expectation value is much smaller than the string scale.
For example, if $gv_0 \sim 100 \ {\rm GeV}$, the radius of the revolution is $d \sim gv_0/M_s^2$.
Then the velocity  of the revolving D3-brane is given by
 $\omega d \sim gv_0/M_s \sim 10^{-16}$, and the motion is quite non-relativistic.
The acceleration  is given by $ \omega^2 d\sim gv_0$,
 and it is again much  smaller than the string scale.  
It justifies the validity of the DBI action and its expansions.

\section{Gauge symmetry breaking by revolving D3-branes}
\label{sec:symmetry_breaking}
In this section
 we investigate the low-energy spectrum of the world-volume theory with revolving D3-branes on the third torus,
 and calculate the masses of various fields.
We also discuss stability of the vacuum expectation value.
 
If the centrifugal potential is stronger than the tension,  we can 
treat each point on the brane is bound at the minimum of the potential (\ref{VM})
\footnote{Note that in the opposite situation where the centrifugal potential is 
 comparable to or weaker than the tension of the brane, 
 a different treatment of the effect of revolution is necessary by
 expanding the action of the system around the time-dependent background with only
 the total angular momentum being conserved. In this case, the term linear in
 the time-derivative cannot be removed and the effect of the Lorentz violation on the D3-branes
 appears. Detailed investigations will be studied in a separate paper.}.
In such a situation, the Lagrangian we are going to study
 becomes\footnote{The convention of the signature of metric is different from that used in the DBI action.}
\begin{equation}
 {\cal L} = | D^\mu Z_i^{(a)} |^2 -(V_F  + V_D + V_M) \ .
\label{L-model}
\end{equation}
$V_F$ and $V_D$ are the F and D-term potentials. 
The potential $V_M$
 is the supersymmetry breaking potential induced by the 
 anti-D7-branes and the revolution of D3-branes discussed in the previous section.
This term determines the position of D3-branes revolving around the fixed point.
The solution of the stationary condition of the potential
 is given by eq.~(\ref{Z-sol-revolv}) with $v_0^2=l/\mu$,
 and then the gauge symmetry is spontaneously broken
 from U$(2) \times$U$(1)_1 \times$U$(1)_2$ to U$(1) \times$U$(1)$.

\subsection{Gauge boson masses}
First let us calculate the masses of the gauge bosons.
Since the gauge symmetry breaking is from 
U$(2) \times$U$(1)_1 \times$U$(1)_2$  to U$(1) \times$U$(1)$, 
four of the six gauge bosons,
 $W^A_\mu$ of U$(2)$, $B^{(1)}_\mu$ of U$(1)_1$ and $B^{(2)}_\mu$ of U$(1)_2$,
 become massive by the Higgs mechanism.
The other two gauge fields, associated with two unbroken generators,  remain massless.
With the normalizations of the charges in the covariant derivatives of eq.~(\ref{Cov-Der}),
 the following two combinations 
\begin{equation}
 A^{(1)}_\mu = {1 \over \sqrt{2}} W^3_\mu - {1 \over 2} \left( B^{(1)}_\mu + B^{(2)}_\mu \right),
\qquad
 A^{(2)}_\mu = \sqrt{2 \over 3} W^0_\mu + {1 \over \sqrt{6}} W^3_\mu
             + {1 \over {2\sqrt{3}}} \left( B^{(1)}_\mu + B^{(2)}_\mu \right).
\label{unbroken-gauge}
\end{equation}
 are massless. 
The other four combinations become massive.  
They are classified into two different types.
The first type is associated with the broken generators of U$(2)$ and given by
\begin{equation}
 W^{\pm}_\mu = {1 \over \sqrt{2}} \left( W^1_\mu \pm i W^2_\mu \right) .
\end{equation}
They acquire mass of $gv_0$ and 
 are charged under the above unbroken U$(1)$ symmetries\footnote{It is neutral under
 the combination of $(A_\mu^{(1)}-\sqrt{3} A_\mu^{(2)})/2$.}.
The other type of combinations is orthogonal to the massless gauge fields
 (\ref{unbroken-gauge}) and written as 
\begin{equation}
 Z^{(1)}_\mu = {1 \over \sqrt{2}} \left( B^{(1)}_\mu - B^{(2)}_\mu \right),
\qquad
 Z^{(2)}_\mu = {1 \over \sqrt{3}} \left( W^0_\mu - W^3_\mu \right)
             - {1 \over \sqrt{6}} \left( B^{(1)}_\mu + B^{(2)}_\mu \right) .
\end{equation}
They acquire masses of $\sqrt{3} gv_0$ and
 are neutral under the remaining two U$(1)$ symmetries.
The gauge boson $Z^{(1)}_\mu$ corresponds to an anomalous U$(1)$ gauge symmetry of U$(1)_1-$U$(2)_2$.
These two values of masses
 can be geometrically understood as the distances of  the stretched open strings
 between the D3-branes.
Namely,
 $gv$ corresponds to the distance $d$ of the D3-branes from the center
 and $\sqrt{3}gv_0$ corresponds to the distance $s$ between the D3-branes in Figure~\ref{fig:separation}. 

\subsection{Scalar boson masses}
Next let us calculate the mass spectrum of the scalar fields.
As we discussed at the end of the previous section, 
 since the potential $V_M$ is a function of a single combination of fields,
 only the breathing mode of eq.~(\ref{breathing-mode}) acquires mass through
 the potential $V_M$ and the other modes are not affected by the
 supersymmetry breaking potential $V_M$.
We now consider the fluctuations of the scalar fields along the third torus $Z_3^{(a)}$.
Define the fluctuations as 
 $Z_3^{(a)} = \langle Z_3^{(a)} \rangle + \varphi^a$,
 where the vacuum expectation value is defined in eq.~(\ref{Z-sol-revolv}) and $\varphi^a$ are decomposed into three types of scalar fields $\sigma, \rho$ and $\pi$;
\begin{equation}
 \varphi^1 =
  \left( \begin{array}{c} \rho^1 \\ (\sigma^1 + i \pi^1)/\sqrt{2} \end{array} \right),
\qquad
 \varphi^2 =
  \left( \begin{array}{c} \rho^2 \\ (\sigma^2 + i \pi^2)/\sqrt{2} \end{array} \right),
\qquad
 \varphi^3 = {1 \over \sqrt{2}} (\sigma^3 + i \pi^3).
\end{equation}
The mass terms of these scalars can be obtained
 by expanding the potential $V_F+V_D+V_M$ around the classical solution of eq.~(\ref{Z-sol-revolv}).
The F-term potential $V_F$ does not generate any mass term,
 because the classical solution is non-vanishing only along the third direction $Z_3^{(a)}$. 
The potential $V_M$ generates the mass term for the breathing mode in $\sigma$ only,
 and the other modes of $\sigma$, $\pi$ and $\rho$ are not affected by $V_M$.

The mass matrix of the $\sigma$ modes are obtained as follows.
By redefining the fluctuations of $\sigma^a$ by  the unitary transformation,
\begin{equation}
 \left(
  \begin{array}{c}
   {\tilde \sigma}^1 \\ {\tilde \sigma}^2 \\ {\tilde \sigma}^3
  \end{array}
 \right)
 = U
 \left(
  \begin{array}{c}
   \sigma^1 \\ \sigma^2 \\ \sigma^3
  \end{array}
 \right),
\qquad
 U = \left(
      \begin{array}{ccc}
       1/\sqrt{3} & 1/\sqrt{3} & 1/\sqrt{3} \\
       1/\sqrt{2} & -1/\sqrt{2} & 0 \\
       1/\sqrt{6} & 1/\sqrt{6} & -\sqrt{2/3} \\
      \end{array}
     \right),
\end{equation}
 only the first component $\tilde{\sigma}^1$, which corresponds to the breathing mode,
 acquires mass proportional to $\mu$.
Since this mode originally corresponds to the flat direction of the D-term potential,
 $V_D$ does not generate potential for the breathing mode.
On the other hand,
 the other two combinations get their masses from the D-term potential only, and not affected by $V_M$.
Their masses are proportional to $gv_0$, and independent of $\mu \sim M_s/10$.
Indeed, the mass matrix of ``Higgs bosons'', $\sigma^a$, is given by
\begin{equation}
 {1 \over 3} (2 \mu)^2
  \left(
   \begin{array}{ccc}
    1 + 2 \epsilon & 1-\epsilon & 1-\epsilon \\
    1-\epsilon & 1 + 2 \epsilon & 1-\epsilon \\
    1-\epsilon & 1-\epsilon & 1 + 2 \epsilon
   \end{array} 
  \right),
\end{equation}
 where $\epsilon \equiv (gv_0/2 \mu)^2/3$.
The mass of the breathing mode $\tilde{\sigma}^1$ is  $2\mu$ and becomes very heavy.
The other two modes $\tilde{\sigma}^2$ and $\tilde{\sigma}^3$ have equal masses of
$\sqrt{3} g v_0$.

The masses of the $\rho$ modes are given as follows.
Among four real scalar fields in $\rho^a$,
 two combinations are would-be Nambu-Goldstone bosons of the gauge symmetry breaking of SU$(2)$ into U$(1)$.
Since the broken generators acting on $Z_3^{(1)}$ and $Z_3^{(2)}$ are
 $\sigma^a/2$ and $(-\sigma^{a})^*/2$ with $a=1,2$ respectively,
the combinations of
\begin{equation}
 {1 \over \sqrt{2}} ( \rho^1_R + \rho^2_R ),
\qquad
 {1 \over \sqrt{2}} ( \rho^1_I - \rho^2_I )
\end{equation}
 are eaten by the longitudinal modes of  $W^\pm_\mu$ gauge bosons.
Here, we wrote the real and imaginary part of $\rho^a$ as
 $\rho^a = (\rho^a_R + i \rho^a_I)/\sqrt{2}$.
The other two combinations of
\begin{equation}
 {1 \over \sqrt{2}} ( \rho^1_R - \rho^2_R ),
\qquad
 {1 \over \sqrt{2}} ( \rho^1_I + \rho^2_I )
\end{equation}
 are ``charged Higgs bosons'' with mass $gv_0$,

For the  $\pi^a$ modes,
 two of them are the would-be Nambu-Goldstone bosons of the gauge symmetry breaking of
 four U$(1)$ symmetries (corresponding to two diagonal generators of U$(2)$ and U$(1)_1$ and U$(1)_2$)
 into U$(1) \times$U$(1)$,
 and become the longitudinal modes of $Z^{(1)}_\mu$ and $Z^{(2)}_\mu$.
The last one is massless.
It is the Nambu-Goldstone boson
 associated with the breaking of the global U$(1)$ symmetry,
 by which the phases of the scalar fields $Z_3^{(a)}$ $(a=1,2,3)$ are rotated simultaneously
 as $Z_3^{(a)} \rightarrow e^{i \theta} Z_3^{(a)}$.
This mode corresponds to changing the positions of revolving D3-branes
 without changing their relative positions and the distance from the origin.
This global symmetry is the anomalous U$(1)_R$ symmetry
 of the full theory with the superpotential of eq.~(\ref{superpotential}),
 and accordingly the corresponding scalar field becomes massive.

\subsection{Stability of the symmetry breaking scale}
The following arguments indicate stability of the mass spectrum of the scalar fields against radiative corrections. 
The spontaneous symmetry breaking occurs due to the potential $V_M$ in eq.~(\ref{L-model}).
If gauge interactions were absent,
 we could consider the upper and the lower components of the doublet fields $Z_3^{(a)}$ with $a=1,2$ 
 as independent complex scalar fields.
The action is then invariant under the $U(5)$ rotation of 5 complex scalar fields,
 and the $U(5)$ invariant potential $V_M$ has a minimum where U$(5)$ is spontaneously broken down to U$(4)$.
Then the symmetry breaking of the global U$(5)$ symmetry produces
 one massive ``Higgs'' boson and nine massless Nambu-Goldstone bosons.
In presence of the gauge interactions, 
 four of these nine Nambu-Goldstone bosons are absorbed into the gauge bosons through the Higgs mechanism,
 and the remaining five scalars become massive as pseudo-Nambu-Goldstone bosons (pNGBs),
 because the corresponding U$(5)$ symmetries are explicitly broken by the gauge interactions.
The radiative correction to the mass of the pNGB
 is protected by the approximate symmetry.
The mass spectrum of the pNGB
 is generally written as the Dashen formula \cite{Dashen:1969eg} 
\begin{equation}
m^2 = {1 \over {f^2}} \langle [Q, [Q , H]] \rangle, 
\end{equation}
 where $f$ is the decay constant and
  $Q$ is the broken generator corresponding to the fluctuation 
of the pNGB  around the minimum,
 and stable under radiative corrections so long as the coupling is weak
 \cite{Weinberg:1972fn,Weinberg:1975gm,ArkaniHamed:2001nc, ArkaniHamed:2002qx}.
In our scenario the vacuum expectation value $v_0$
 is determined in terms of the initial angular momentum of the D3-branes which is conserved. 
Hence, it is fixed by the input of the initial condition, not by the field theory itself.
This guarantees the stability of the value of the distance between the revolving D3-branes
 and the fixed point where anti-D7-branes are localized.
 In this geometrical way the symmetry breaking scale is invariant\footnote{
 In section \ref{sec:conclusion},
 we discuss possible decay of the revolution 
 radius (i.e., the vacuum expectation value) through
  emissions of massless closed string states.}.
 Furthermore, since the mass of the ${\tilde \sigma}_1$ field is much heavier
 than the other scalar fields, quantum fluctuations of the vacuum expectation
 value are highly 
 suppressed  in the low-energy effective theory.
 This is another indication of the stability of the scale 
 of the spontaneous gauge symmetry breaking.
 
\subsection{Other fields and fermions}
So far we have considered only the scalar fields $Z_3^{(a)}$ on the third torus.
Let us now see how the masses for other fields $Z_i^{(a)}$ with $i=1,2$ are generated.
They were originally moduli fields
 corresponding to the motion of D3-branes along the first and second tori. 
 When $Z_3^{(a)}$ have the vacuum expectation values of eq.~(\ref{Z-sol-revolv}),
they acquire their masses
 through the interactions described by the F-term potential (\ref{F-potential}).
For their fermionic superpartners,
 they acquire the same masses through the Yukawa couplings,
 since the gauge symmetry breaking by the vacuum expectation value (\ref{Z-sol-revolv})
 does not break supersymmetries.
All the fields, which are associated with the first and second tori,
 obtain masses of the order of $gv_0$.
Thus all the  
 D-brane moduli fields corresponding to the first and the second tori are stabilized.
They may also obtain their masses through one-loop effects
 of open strings between D3 and anti-D7 branes, which are not supersymmetric.

Yukawa couplings between the scalar fields $Z_i^{(a)}$ and the 
fermion fields  in the chiral multiplets come from the superpotential of eq.~(\ref{superpotential}).
Because of the $\epsilon$-tensor in the superpotential,
 the Yukawa couplings must contain at least one field 
 associated with each of the three tori.
Hence,
since the scalar fields $Z_3^{(a)}$ with the nonzero vacuum expectation values
couple to the fermions associated with the first and the second tori, 
 the fermion in the first and the second tori become massive in pairs.
Especially, the Yukawa coupling of  the ``Higgs boson'', ${\tilde \sigma}_1$,
 with the fermions are proportional to their masses.
 But it is not the case for light scalar fields, ${\tilde \sigma}_2$ and ${\tilde \sigma}_3$.
Note that the fermion fields  associated with the third torus 
 do not obtain their masses through the Yukawa couplings. 
 A brief discussion towards  
 more realistic model constructions of the standard model  is given 
in section~\ref{sec:conclusion}.

\section{Conclusions and discussions}
\label{sec:conclusion}

In this paper, 
 we have proposed a new mechanism of spontaneous gauge symmetry breaking
 in string theory with revolving D-branes around an orbifold fixed point. 

In orbifold compactification of the string theory two types of D-branes are known: 
 ordinary D-branes and fractional D-branes. 
The ordinary D-branes can move away from the fixed points by forming
 an invariant set under the orbifold projections
 while the fractional D-branes are localized at the fixed points.
When the ordinary D-branes move away from the fixed point, 
 the gauge symmetry is spontaneously broken with reduction of the rank of the gauge group.
 If the separation scale of the D-branes is shorter than the string length
 and their relative motion is non-relativistic,
 the dynamics can be well described by the low-energy world-volume theories of D-branes \cite{Douglas:1996yp}.
The separation length corresponds to the vacuum expectation value of the D-brane moduli fields.
Unless supersymmetry is broken, the moduli fields have the flat potential along the 
vacuum expectation values. 
If we consider configurations of D-branes with broken supersymmetries, 
 the flat directions will be lifted up by the one-loop corrections,
 whose mass scales are generically given by the string scale 
  with a small correction of the one-loop suppression factors.
  Such effects are responsible for  attractive forces between D-branes and anti-D-branes.
 
In the present paper, we consider a particular model of D3 and anti-D7 branes
 at a fixed point of a $T^6/{\rm Z}_3$ orbifold.
In order to stabilize the separation of D3-branes from the fixed point, 
 we introduced revolution of D3-branes around  anti-D7-branes localized at the fixed point.
The revolution can be described as the time-dependent phase of the classical solution of D-brane moduli fields. 
In the rest frame of the revolving D3-branes, 
 the centrifugal potential appears. This force can balance the attractive force between the D3 and anti-D7 branes.  
The distance of the revolving D3-branes from the fixed point is determined 
 by the value of the angular momentum, which should be given by
 the initial condition of the cosmological evolution of the D-brane configuration.
An important point is that the scale of the distance, namely the vacuum expectation value of the moduli fields,
 is determined in terms of the angular momentum and can be taken
  much lower than the string scale.
The situation is quite similar to the revolution radius of the earth around the sun,
 where it was determined by the initial condition of the solar system evolution.
We can imagine that the early universe are filled with gas of D-branes and anti-D-branes.
They collide and some of them are annihilated,
 but due to the conservation of the angular momentum, some branes  revolve around others.

We calculated the mass spectrum of various moduli fields.
The breathing mode, in
which all the nonzero expectation values of the moduli fields change simultaneously,
 becomes super heavy as the string scale $M_s$ because the mass comes 
 from the balance of the attractive potential between D3 and $\overline{\rm D7}$
 and the repulsive centrifugal potential.
On the other hand, 
the other scalar modes acquire their masses through the gauge symmetry breaking,
and hence
 their masses are of order of the vacuum expectation value $v_0 \ll M_s$.
They are
 stable under radiative corrections due to their nature of the pseudo-Nambu-Goldstone bosons.  
It is interesting to note that the stability of the scale $v_0$ is related to
the stability of the geometrical distance of the revolving D3 branes from the fixed point.
The distance is classically stable 
 due to the conservation of angular momentum. 

Now several important remarks are in order. 
The first remark is possible
loss of the angular momentum of the revolving D3-branes by emitting massless 
closed string states (gravitons or 4-form R-R field).
If the effects are large, the revolution frequency $\omega$ 
decreases and the D3-branes fall into the fixed point rapidly.
As we discussed at the end of section  \ref{sec:rotating_D3-branes}, 
the velocity $\beta=\omega d \sim gv_0/M_s$ and the acceleration $\alpha=\omega^2 d \sim gv_0$
are very small and the emission may be expected to be tiny. 
However, it is not the case since both of the tension and the R-R charge
are very large. 
Here let us estimate the emission rate of the 
 gravitational radiation into the bulk 
 from the revolving D3-brane by using the formula in \cite{Bachlechner:2013fja}. 
 The energy emission rate per unit time and unit volume of the D3-brane
  is estimated by  $M_s^5 (\omega d)^4$.
 Huge factor comes from the tension  of the D3-brane. 
The rate is very large in comparison with the kinetic energy of the 
revolving D3-branes per unit volume
 $\tau_3 (\omega d)^2/2 \sim M_s^4 (\omega d)^2$,
 even though $\omega d \sim (\omega/M_s)(v/M_s)$ is very small.
If it is really the case, the stability of the vacuum expectation value will be lost.
But, the formula of the emission rate may not be applicable to the present case since the
typical frequency of the radiation, which is of the same order of the
angular velocity of the revolution  $\omega \simeq M_s/10$, 
is close to the cut off scale of the low-energy effective theory, $M_s$. 
Also,  it is not certain whether radiation itself is possible since the length scale of 
 the compact space is comparable to 
 the typical wave length of the emitted radiation.
Furthermore, 
 the NS-NS tadpoles are not cancelled in this non-supersymmetric configurations of
 D-branes and 
we may need to take the effects of the backreaction to the space-time metric, dilaton and B-fields,
 which may drastically change the emission rate of radiation. 
So more detailed analysis
 will be necessary to give the correct estimates of the emission rate.


The second remark is a possibility of a 
 modification of gravity in the IR region.
One may expect that the revolution of D3-branes affects the gravity in
the IR region since our model is similar 
 to the ghost condensation \cite{ArkaniHamed:2003uy}.
In both of the ghost condensation and our setting, 
 the time-dependent classical solutions of the scalar fields play an important role.
In the ghost condensation, $X^2=|\partial_\mu \phi|^2$ has a non-vanishing vacuum
 expectation value at the minimum $X_0^2$ of the function $P(X^2)$.
Then the field can be expanded around 
 the time-dependent solution $\langle \phi \rangle = X_0 t$,
 which breaks the Poincar\'e symmetry.
Writing $\phi = X_0 t +\varphi$, we have $X^2=(X_0 + \dot{\varphi})^2-(\nabla \varphi)^2$.
Then the kinetic term $P(X^2)$ becomes $P(X^2) \sim P(X_0^2)+[2 P''(X_0^2) X_0^2] \ \dot{\varphi}^2$.
The $(\nabla \varphi)^2$ term is absent,
 and the IR dispersion relation is drastically modified as $\omega^2 \varpropto p^4$.
In our case of the revolving D-branes,
 the time-dependent solution appears not because it gives a minimum of the kinetic term 
 but it is due to the centrifugal potential in the rest frame of the D-branes. 
Because of this, if the centrifugal potential is stronger than the 
tension of the brane and each point of the brane is bound strongly at the minimum
of the potential,  
 the spectrum of the fluctuation seems to keep the Lorentz invariance and different 
 from the case of the ghost condensation. But Lorentz invariance is generally broken
 due to the revolution, especially if the tension is comparable to the centrifugal potential.
We come back to this issue in near future.

The third remark is a construction of more realistic models of particle physics. 
There have been many efforts to construct realistic models
 in the system of D-branes at orbifold fixed points or other singular points, in general
 (see for example \cite{Aldazabal:2000sa,Berenstein:2001nk,Verlinde:2005jr,Krippendorf:2010hj}).
Though the aim of this paper is not to propose a realistic model,
 some interesting structures of realistic models are included in the model
 we have discussed above.
For example,
 following \cite{Aldazabal:2000sa},  
 we can identify the subgroup SU$(2)$ in U$(2)$ as the standard model SU$(2)_L$ group
and  a non-anomalous combination of U$(1)$ symmetries,
 $Q_Y=-(Q/2+Q_1+Q_2)$, as the U$(1)_Y$ in the standard model. 
Here $Q$, $Q_1$ and $Q_2$ are charges of U$(1)$ in U$(2)$, U$(1)_1$ and U$(1)_2$, respectively.
In the identification,  the value of the Weinberg angle becomes $\sin^2 \theta_W \simeq 0.27$.
Since the charges are normalized to take values of $\pm1$,
 we can identify the scalar component of $Z^{(1)}_3$ as a Higgs doublet field ($Q_Y = +1/2$),
 the fermion components of $Z^{(2)}_{i=1,2}$ as the left-handed lepton doublets of the first and 
 the second generations  ($Q_Y = -1/2$),
 and the fermion components of $Z^{(3)}_{i=2,1}$ as the right-handed neutrinos of the first 
 and the second generations ($Q_Y = 0$).
In this simple setting, the Yukawa couplings are originated from 
the superpotential of eq.~(\ref{superpotential}) and give large masses to neutrinos. 
It would be interesting to try to construct more realistic models with quarks and leptons
 in the framework of the spontaneous gauge symmetry breaking by the revolution of D-branes. 
 
The fourth remark
 is a realization of an inflationary universe with the revolving D-branes. 
In the present paper, we treated the bulk space-time as the flat Minkowski,
 but of course it receives huge backreaction effects, since the configuration
 is not supersymmetric.
In the following, 
 instead of solving the bulk 10-dimensional Einstein equation,
 we study time-evolution of the induced metric on the D3-branes \cite{Kehagias:1999vr}. 
Since D3-branes are isotropic and homogeneous, 
the time-evolution of the metric will be described by the FRW universe\footnote{
Our assumption here is the metrics on different D3-branes are strongly correlated 
each other 
so that they can be described by a single common metric.}
with the scale factor $a(t)$.
In the FRW universe, 
the solution to the equation
of motion is changed from eq.~(\ref{classical-solution-massivescalar}) to 
\begin{equation}
 \phi \sim a^{-3/2} v_0 e^{i \omega t}.
\end{equation}
 This solution is valid
 when the oscillation frequency $\omega=\mu$ is larger than the Hubble parameter
 $H=\dot{a}/a$.
In the opposite case $H > \mu$, the field is frozen 
  so that the prefactor is almost constant. 
Such a scale factor dependence is understandable
 since the oscillation in a harmonic potential behaves as non-relativistic particles
 and the energy density $\mu^2 v^2$ must decay as $a^{-3}$. 
Now Suppose that D3-branes, while they are rotating,
 start rolling down along the potential from a larger value of $v$ to the stabilized point $v_0$.
 The current $j_0$ is  proportional to  $a^{-3} v^2 \omega$. 
 Due to an additional factor $\sqrt{|g|}=a^3$ in the current conservation
 $\nabla^\mu j_\mu = a^{-3} \partial^\mu (a^3 j_\mu) =0 $,  
 the same quantity $l=v^2 \omega$ is conserved. 
The Hamiltonian density  $\rho$  for $\mu > H$ becomes
\begin{equation}
\rho = a^{-3} (v^2 \omega^2 + \mu^2 v^2)=a^{-3} (l^2/v^2 +\mu^2 v^2) .
\end{equation}
The minimum of the potential is given by the same value $v_0=l/\mu$.
The pressure density $p$ is given by changing the sign of the second term as
\begin{equation}
 p = a^{-3} (l^2/v^2 -\mu^2 v^2) .
\end{equation}
This is because
 the first term (centrifugal potential) is originated from the kinetic energy 
 while the second one comes from the attractive interaction between D3 and anti-D7 branes.
Denoting the ratio of $v$ to $v_0=\sqrt{l/\mu_0}$ by $v/v_0 \equiv e^{\theta/2}$,
 $\rho$ and $p$ can be written as $\rho = 2 a^{-3} l \mu \cosh \theta$ and $p=- 2 a^{-3} l \mu \sinh \theta$.
Then the equation of state becomes
\begin{equation}
 w = \frac{p}{\rho} = - \tanh \theta .
\end{equation}
It is interesting that
 the equation of state of the revolving motion interpolates between $w=-1$ and $w=0$.
At the final stage of the evolution, D3-branes are stabilized at $v=v_0$
 and the equation of state is $w=0$.
The energy density is then given by $\rho =2 l \mu a^{-3}$.
On the other hand, in the initial stage of the cosmic evolution,
 the second term $\mu^2 v^2$ in $\rho$ and $p$ dominates the first term and 
 the equation of state is $w=-1$.
In this case, the field is frozen and the prefactor $a^{-3}$ is replaced by a constant. 
Then the energy density becomes $\rho = \mu^2 v^2$,
 which gives the Hubble constant of the inflationary universe caused by the revolving D3-branes.
In this scenario,
 the radial direction of the revolving D-branes, $\tilde{\sigma}_1$,
 plays a role of the inflaton field, similar to the brane-inflation model \cite{Dvali:1998pa}.
Similar angular motions of D-branes in the DBI inflation scenario
 has been discussed in \cite{Easson:2007fz,Baumann:2007ah,Easson:2007dh}.

The final remark is about possible modifications of the background geometry of the compact space
due to introductions of the three-form fluxes and additional (anti)-D-branes 
for the moduli stabilization.
The form of the angular momentum conservation might be modified as well as the attractive potential. But as far as the rotational symmetry is preserved and the angular momentum
is conserved, the essential mechanism will not be changed much. 
The non-trivial modification of the background geometry
could make it easier  to obtain the small scale of the gauge symmetry breaking
 in comparison to the string scale, as 
is the case in the warped geometries \cite{Cascales:2003wn}.

Each of the above mentioned issues is very interesting, 
 but needs more careful and detailed investigations.
We want to come back to these issues in future.

\section*{Acknowledgments}

The authors would like to thank Ryuichiro Kitano and Yutaka Sakamura
 for discussions and useful comments.
This work was supported in part by Grant-in-Aid for Scientific Research
 (\# 23540329, \# 23244057 and \# 26400253) from MEXT Japan.
NK would like to thank KEK for the kind hospitality.


\begin{thebibliography}{99}

\bibitem{Weinberg:1979bn}
  S.~Weinberg,
  ``Implications of Dynamical Symmetry Breaking: An Addendum,''
  Phys.\ Rev.\ D {\bf 19} (1979) 1277.
\bibitem{Susskind:1978ms}
  L.~Susskind,
  ``Dynamics of Spontaneous Symmetry Breaking in the Weinberg-Salam Theory,''
  Phys.\ Rev.\ D {\bf 20} (1979) 2619.
\bibitem{Ibanez:1982fr}
  L.~E.~Ibanez and G.~G.~Ross,
  ``SU(2)-L x U(1) Symmetry Breaking as a Radiative Effect of Supersymmetry Breaking in Guts,''
  Phys.\ Lett.\ B {\bf 110} (1982) 215.
\bibitem{Inoue:1982pi}
  K.~Inoue, A.~Kakuto, H.~Komatsu and S.~Takeshita,
  ``Aspects of Grand Unified Models with Softly Broken Supersymmetry,''
  Prog.\ Theor.\ Phys.\  {\bf 68} (1982) 927
   [Prog.\ Theor.\ Phys.\  {\bf 70} (1983) 330].
\bibitem{Coleman:1973jx}
  S.~R.~Coleman and E.~J.~Weinberg,
  ``Radiative Corrections as the Origin of Spontaneous Symmetry Breaking,''
  Phys.\ Rev.\ D {\bf 7} (1973) 1888.
\bibitem{Meissner:2006zh}
  K.~A.~Meissner and H.~Nicolai,
  ``Conformal Symmetry and the Standard Model,''
  Phys.\ Lett.\ B {\bf 648} (2007) 312
  [hep-th/0612165].
\bibitem{Foot:2007as}
  R.~Foot, A.~Kobakhidze and R.~R.~Volkas,
  ``Electroweak Higgs as a pseudo-Goldstone boson of broken scale invariance,''
  Phys.\ Lett.\ B {\bf 655} (2007) 156
  [arXiv:0704.1165 [hep-ph]].
\bibitem{Iso:2009ss}
  S.~Iso, N.~Okada and Y.~Orikasa,
  ``Classically conformal $B^-$ L extended Standard Model,''
  Phys.\ Lett.\ B {\bf 676} (2009) 81
  [arXiv:0902.4050 [hep-ph]].
\bibitem{Iso:2009nw}
  S.~Iso, N.~Okada and Y.~Orikasa,
  ``The minimal B-L model naturally realized at TeV scale,''
  Phys.\ Rev.\ D {\bf 80} (2009) 115007
  [arXiv:0909.0128 [hep-ph]].
\bibitem{Iso:2012jn}
  S.~Iso and Y.~Orikasa,
  ``TeV Scale B-L model with a flat Higgs potential at the Planck scale - in view of the hierarchy problem -,''
  PTEP {\bf 2013} (2013) 023B08
  [arXiv:1210.2848 [hep-ph]].
\bibitem{Hosotani:1983xw}
  Y.~Hosotani,
  ``Dynamical Mass Generation by Compact Extra Dimensions,''
  Phys.\ Lett.\ B {\bf 126} (1983) 309.
\bibitem{Cremades:2003qj}
  D.~Cremades, L.~E.~Ibanez and F.~Marchesano,
  ``Yukawa couplings in intersecting D-brane models,''
  JHEP {\bf 0307} (2003) 038
  [hep-th/0302105].
\bibitem{Cvetic:2003ch}
  M.~Cvetic and I.~Papadimitriou,
  ``Conformal field theory couplings for intersecting D-branes on orientifolds,''
  Phys.\ Rev.\ D {\bf 68} (2003) 046001
   [Phys.\ Rev.\ D {\bf 70} (2004) 029903]
  [hep-th/0303083].
\bibitem{Abel:2003vv}
  S.~A.~Abel and A.~W.~Owen,
  ``Interactions in intersecting brane models,''
  Nucl.\ Phys.\ B {\bf 663} (2003) 197
  [hep-th/0303124].
\bibitem{Kitazawa:2004hz}
  N.~Kitazawa,
  ``Dynamical generation of Yukawa interactions in intersecting D-brane models,''
  JHEP {\bf 0411} (2004) 044
  [hep-th/0403278].
\bibitem{Kitazawa:2004nf}
  N.~Kitazawa, T.~Kobayashi, N.~Maru and N.~Okada,
  ``Yukawa coupling structure in intersecting D-brane models,''
  Eur.\ Phys.\ J.\ C {\bf 40} (2005) 579
  [hep-th/0406115].
\bibitem{Blumenhagen:2006xt}
  R.~Blumenhagen, M.~Cvetic and T.~Weigand,
  ``Spacetime instanton corrections in 4D string vacua: The Seesaw mechanism for D-Brane models,''
  Nucl.\ Phys.\ B {\bf 771} (2007) 113
  [hep-th/0609191].
\bibitem{Aldazabal:2000sa}
  G.~Aldazabal, L.~E.~Ibanez, F.~Quevedo and A.~M.~Uranga,
  ``D-branes at singularities: A Bottom up approach to the string embedding of the standard model,''
  JHEP {\bf 0008} (2000) 002
  [hep-th/0005067].
\bibitem{Berenstein:2001nk}
  D.~Berenstein, V.~Jejjala and R.~G.~Leigh,
  ``The Standard model on a D-brane,''
  Phys.\ Rev.\ Lett.\  {\bf 88} (2002) 071602
  [hep-ph/0105042].
\bibitem{Verlinde:2005jr}
  H.~Verlinde and M.~Wijnholt,
  ``Building the standard model on a D3-brane,''
  JHEP {\bf 0701} (2007) 106
  [hep-th/0508089].
\bibitem{Krippendorf:2010hj}
  S.~Krippendorf, M.~J.~Dolan, A.~Maharana and F.~Quevedo,
  ``D-branes at Toric Singularities: Model Building, Yukawa Couplings and Flavour Physics,''
  JHEP {\bf 1006} (2010) 092
  [arXiv:1002.1790 [hep-th]].
\bibitem{Sugimoto:1999tx}
  S.~Sugimoto,
  ``Anomaly cancellations in type I D-9 - anti-D-9 system and the USp(32) string theory,''
  Prog.\ Theor.\ Phys.\  {\bf 102} (1999) 685
  [hep-th/9905159].
\bibitem{Antoniadis:1999xk}
  I.~Antoniadis, E.~Dudas and A.~Sagnotti,
  ``Brane supersymmetry breaking,''
  Phys.\ Lett.\ B {\bf 464} (1999) 38
  [hep-th/9908023].
\bibitem{Angelantonj:1999jh}
  C.~Angelantonj,
  ``Comments on open string orbifolds with a nonvanishing B(ab),''
  Nucl.\ Phys.\ B {\bf 566} (2000) 126
  [hep-th/9908064].
\bibitem{Aldazabal:1999jr}
  G.~Aldazabal and A.~M.~Uranga,
  ``Tachyon free nonsupersymmetric type IIB orientifolds via Brane - anti-brane systems,''
  JHEP {\bf 9910} (1999) 024
  [hep-th/9908072].
\bibitem{Angelantonj:1999ms}
  C.~Angelantonj, I.~Antoniadis, G.~D'Appollonio, E.~Dudas and A.~Sagnotti,
  ``Type I vacua with brane supersymmetry breaking,''
  Nucl.\ Phys.\ B {\bf 572} (2000) 36
  [hep-th/9911081].
\bibitem{Kitazawa:2012hr}
  N.~Kitazawa and S.~Kobayashi,
  ``Spontaneous Gauge Symmetry Breaking in a Non-Supersymmetric D-brane Model,''
  Phys.\ Lett.\ B {\bf 720} (2013) 373
  [arXiv:1211.1777 [hep-th]].
 
\bibitem{Aldazabal:1998mr}
  G.~Aldazabal, A.~Font, L.~E.~Ibanez and G.~Violero,
  ``D = 4, N=1, type IIB orientifolds,''
  Nucl.\ Phys.\ B {\bf 536} (1998) 29
  [hep-th/9804026].
\bibitem{Kachru:2003aw}
  S.~Kachru, R.~Kallosh, A.~D.~Linde and S.~P.~Trivedi,
  ``De Sitter vacua in string theory,''
  Phys.\ Rev.\ D {\bf 68} (2003) 046005
  [hep-th/0301240].
\bibitem{Kachru:2003sx}
  S.~Kachru, R.~Kallosh, A.~D.~Linde, J.~M.~Maldacena, L.~P.~McAllister and S.~P.~Trivedi,
  ``Towards inflation in string theory,''
  JCAP {\bf 0310} (2003) 013
  [hep-th/0308055].
\bibitem{Antoniadis:2004pp}
  I.~Antoniadis and T.~Maillard,
  ``Moduli stabilization from magnetic fluxes in type I string theory,''
  Nucl.\ Phys.\ B {\bf 716} (2005) 3
  [hep-th/0412008].
\bibitem{Antoniadis:2006eu}
  I.~Antoniadis, A.~Kumar and T.~Maillard,
  ``Magnetic fluxes and moduli stabilization,''
  Nucl.\ Phys.\ B {\bf 767} (2007) 139
  [hep-th/0610246].
\bibitem{Kitazawa:2014hya}
  N.~Kitazawa,
  ``Towards the stabilization of extra dimensions by brane dynamics,''
  Int.\ J.\ Mod.\ Phys.\ A {\bf 30} (2015) 0055
  [arXiv:1409.8401 [hep-th]].
\bibitem{Fischler:1986ci}
  W.~Fischler and L.~Susskind,
  ``Dilaton Tadpoles, String Condensates and Scale Invariance,''
  Phys.\ Lett.\ B {\bf 171} (1986) 383.
\bibitem{Fischler:1986tb}
  W.~Fischler and L.~Susskind,
  ``Dilaton Tadpoles, String Condensates and Scale Invariance. 2.,''
  Phys.\ Lett.\ B {\bf 173} (1986) 262.
\bibitem{Ibanez:1998qp}
  L.~E.~Ibanez, R.~Rabadan and A.~M.~Uranga,
  ``Anomalous U(1)'s in type I and type IIB D = 4, N=1 string vacua,''
  Nucl.\ Phys.\ B {\bf 542} (1999) 112
  [hep-th/9808139].

\bibitem{Dvali:1998pa}
  G.~R.~Dvali and S.~H.~H.~Tye,
  ``Brane inflation,''
  Phys.\ Lett.\ B {\bf 450} (1999) 72
  [hep-ph/9812483].
\bibitem{ArkaniHamed:2003uy}
  N.~Arkani-Hamed, H.~C.~Cheng, M.~A.~Luty and S.~Mukohyama,
  ``Ghost condensation and a consistent infrared modification of gravity,''
  JHEP {\bf 0405} (2004) 074
  [hep-th/0312099].
 
\bibitem{Dashen:1969eg}
  R.~F.~Dashen,
  ``Chiral SU(3) x SU(3) as a symmetry of the strong interactions,''
  Phys.\ Rev.\  {\bf 183} (1969) 1245.
\bibitem{Weinberg:1972fn}
  S.~Weinberg,
  ``Approximate symmetries and pseudoGoldstone bosons,''
  Phys.\ Rev.\ Lett.\  {\bf 29} (1972) 1698.
\bibitem{Weinberg:1975gm}
  S.~Weinberg,
  ``Implications of Dynamical Symmetry Breaking,''
  Phys.\ Rev.\ D {\bf 13} (1976) 974.
\bibitem{ArkaniHamed:2001nc}
  N.~Arkani-Hamed, A.~G.~Cohen and H.~Georgi,
  ``Electroweak symmetry breaking from dimensional deconstruction,''
  Phys.\ Lett.\ B {\bf 513} (2001) 232
  [hep-ph/0105239].
\bibitem{ArkaniHamed:2002qx}
  N.~Arkani-Hamed, A.~G.~Cohen, E.~Katz, A.~E.~Nelson, T.~Gregoire and J.~G.~Wacker,
  ``The Minimal moose for a little Higgs,''
  JHEP {\bf 0208} (2002) 021
  [hep-ph/0206020].
   
\bibitem{Douglas:1996yp}
  M.~R.~Douglas, D.~N.~Kabat, P.~Pouliot and S.~H.~Shenker,
  ``D-branes and short distances in string theory,''
  Nucl.\ Phys.\ B {\bf 485} (1997) 85
  [hep-th/9608024].
\bibitem{Bachlechner:2013fja}
  T.~C.~Bachlechner and L.~McAllister,
  ``D-brane Bremsstrahlung,''
  JHEP {\bf 1310} (2013) 022
  [arXiv:1306.0003 [hep-th]].
\bibitem{Kehagias:1999vr} 
  A.~Kehagias and E.~Kiritsis,
  ``Mirage cosmology,''
  JHEP {\bf 9911}, 022 (1999)
  [hep-th/9910174].
\bibitem{Easson:2007fz}
  D.~A.~Easson, R.~Gregory, G.~Tasinato and I.~Zavala,
  ``Cycling in the Throat,''
  JHEP {\bf 0704} (2007) 026
  [hep-th/0701252 [HEP-TH]].
\bibitem{Baumann:2007ah}
  D.~Baumann, A.~Dymarsky, I.~R.~Klebanov and L.~McAllister,
  ``Towards an Explicit Model of D-brane Inflation,''
  JCAP {\bf 0801} (2008) 024
  [arXiv:0706.0360 [hep-th]].
\bibitem{Easson:2007dh}
  D.~A.~Easson, R.~Gregory, D.~F.~Mota, G.~Tasinato and I.~Zavala,
  ``Spinflation,''
  JCAP {\bf 0802} (2008) 010
  [arXiv:0709.2666 [hep-th]].
\bibitem{Cascales:2003wn}
  J.~F.~G.~Cascales, M.~P.~Garcia del Moral, F.~Quevedo and A.~M.~Uranga,
  ``Realistic D-brane models on warped throats: Fluxes, hierarchies and moduli stabilization,''
  JHEP {\bf 0402} (2004) 031
  [hep-th/0312051].

\end{thebibliography}
\end{document}